



 \font\twelverm=cmr10 scaled 1200    \font\twelvei=cmmi10 scaled 1200
 \font\twelvesy=cmsy10 scaled 1200   \font\twelveex=cmex10 scaled 1200
 \font\twelvebf=cmbx10 scaled 1200   \font\twelvesl=cmsl10 scaled 1200
 \font\twelvett=cmtt10 scaled 1200   \font\twelveit=cmti10 scaled 1200

 \skewchar\twelvei='177   \skewchar\twelvesy='60


 \def\twelvepoint{\normalbaselineskip=12.4pt
   \abovedisplayskip 12.4pt plus 3pt minus 9pt
   \belowdisplayskip 12.4pt plus 3pt minus 9pt
   \abovedisplayshortskip 0pt plus 3pt
   \belowdisplayshortskip 7.2pt plus 3pt minus 4pt
   \smallskipamount=3.6pt plus1.2pt minus1.2pt
   \medskipamount=7.2pt plus2.4pt minus2.4pt
   \bigskipamount=14.4pt plus4.8pt minus4.8pt
   \def\rm{\fam0\twelverm}          \def\it{\fam\itfam\twelveit}%
   \def\sl{\fam\slfam\twelvesl}     \def\bf{\fam\bffam\twelvebf}%
   \def\mit{\fam 1}                 \def\cal{\fam 2}%
   \def\tt{\twelvett}
   \def\nullspace{\nulldelimiterspace=0pt \mathsurround=0pt }
   \def\big##1{{\hbox{$\left##1\vbox to 10.2pt{}\right.\nullspace$}}}
   \def\Big##1{{\hbox{$\left##1\vbox to 13.8pt{}\right.\nullspace$}}}
   \def\bigg##1{{\hbox{$\left##1\vbox to 17.4pt{}\right.\nullspace$}}}
   \def\Bigg##1{{\hbox{$\left##1\vbox to 21.0pt{}\right.\nullspace$}}}
   \textfont0=\twelverm   \scriptfont0=\tenrm   \scriptscriptfont0=\sevenrm
   \textfont1=\twelvei    \scriptfont1=\teni    \scriptscriptfont1=\seveni
   \textfont2=\twelvesy   \scriptfont2=\tensy   \scriptscriptfont2=\sevensy
   \textfont3=\twelveex   \scriptfont3=\twelveex  \scriptscriptfont3=\twelveex
   \textfont\itfam=\twelveit
   \textfont\slfam=\twelvesl
   \textfont\bffam=\twelvebf \scriptfont\bffam=\tenbf
   \scriptscriptfont\bffam=\sevenbf
   \normalbaselines\rm}


 \def\tenpoint{\normalbaselineskip=12pt
   \abovedisplayskip 12pt plus 3pt minus 9pt
   \belowdisplayskip 12pt plus 3pt minus 9pt
   \abovedisplayshortskip 0pt plus 3pt
   \belowdisplayshortskip 7pt plus 3pt minus 4pt
   \smallskipamount=3pt plus1pt minus1pt
   \medskipamount=6pt plus2pt minus2pt
   \bigskipamount=12pt plus4pt minus4pt
   \def\rm{\fam0\tenrm}          \def\it{\fam\itfam\tenit}%
   \def\sl{\fam\slfam\tensl}     \def\bf{\fam\bffam\tenbf}%
   \def\smc{\tensmc}             \def\mit{\fam 1}%
   \def\cal{\fam 2}%
   \textfont0=\tenrm   \scriptfont0=\sevenrm   \scriptscriptfont0=\fiverm
   \textfont1=\teni    \scriptfont1=\seveni    \scriptscriptfont1=\fivei
   \textfont2=\tensy   \scriptfont2=\sevensy   \scriptscriptfont2=\fivesy
   \textfont3=\tenex   \scriptfont3=\tenex     \scriptscriptfont3=\tenex
   \textfont\itfam=\tenit
   \textfont\slfam=\tensl
   \textfont\bffam=\tenbf \scriptfont\bffam=\sevenbf
   \scriptscriptfont\bffam=\fivebf
   \normalbaselines\rm}


 \def\beginlinemode{\endmode
   \begingroup\parskip=0pt \obeylines\def\\{\par}\def\endmode{\par\endgroup}}
 \def\beginparmode{\endmode
   \begingroup \def\endmode{\par\endgroup}}
 \let\endmode=\par
 {\obeylines\gdef\
 {}}
 \def\singlespace{\baselineskip=\normalbaselineskip}
 
 \def\oneandahalfspace{\baselineskip=\normalbaselineskip
   \multiply\baselineskip by 3 \divide\baselineskip by 2}
 \def\doublespace{\baselineskip=\normalbaselineskip \multiply\baselineskip by
2}
 
 \newcount\firstpageno
 \firstpageno=2

\footline={\ifnum\pageno<\firstpageno{\hfil}\else{\hfil\twelverm\folio\hfil}\fi}

 \let\rawfootnote=\footnote              
 \def\footnote#1#2{{\rm\singlespace\parindent=0pt\rawfootnote{#1}{#2}}}
 \def\raggedcenter{\leftskip=4em plus 12em \rightskip=\leftskip
   \parindent=0pt \parfillskip=0pt \spaceskip=.3333em \xspaceskip=.5em
   \pretolerance=9999 \tolerance=9999
   \hyphenpenalty=9999 \exhyphenpenalty=9999 }
 \def\dateline{\rightline{\ifcase\month\or
   January\or February\or March\or April\or May\or June\or
   July\or August\or September\or October\or November\or December\fi
   \space\number\year}}
 \def\received{\vskip 3pt plus 0.2fill
 \centerline{\sl (Received\space\ifcase\month\or
   January\or February\or March\or April\or May\or June\or
   July\or August\or September\or October\or November\or December\fi
   \qquad, \number\year)}}


 \hsize=6.5truein
 \vsize=8.75truein
 \parskip=\medskipamount
 \twelvepoint            
 \doublespace            
 \overfullrule=0pt       



 \def\title                      
   {\let\bf=\bigtenrm2\null\vskip 3pt plus 0.2fill
    \beginlinemode \doublespace \raggedcenter \bf}

 \def\author                     
   {\vskip 3pt plus 0.2fill \beginlinemode
    \singlespace \raggedcenter}

 \def\affil                      
   {\vskip 3pt plus 0.1fill \beginlinemode
    \singlespace\raggedcenter \sl}

 \def\abstract                   
   {\vskip 3pt plus 0.3fill \beginparmode
    \singlespace\narrower ABSTRACT: }

 \def\endtitlepage               
   {\endpage                     
    \body}

 \def\body                       
   {\beginparmode}               

 \def\head#1{                    
   \vskip 0.5truein     
   {\immediate\write16{#1}
    \raggedcenter \uppercase{#1}\par}
    \nobreak\vskip 0.25truein\nobreak}

 \def\subhead#1{                 
   \vskip 0.25truein             
   {\raggedcenter #1 \par}
    \nobreak\vskip 0.25truein\nobreak}

 \def\refto#1{$^{#1}$}           

 \def\references                 
   {\head{References}            

    \beginparmode
    \frenchspacing \parindent=0pt \leftskip=1truecm
    \parskip=8pt plus 3pt \everypar{\hangindent=\parindent}}

 \gdef\refis#1{\indent\hbox to 0pt{\hss#1.~}}    

 \gdef\journal#1, #2, #3, 1#4#5#6{               

     {\sl #1~}{\bf #2}, #3, (1#4#5#6)}           

 \gdef\journ2 #1, #2, #3, 1#4#5#6{               

     {\sl #1~}{\bf #2}: #3, (1#4#5#6)}           

 \def\refstylenp{                
   \gdef\refto##1{ (##1)}                                
   \gdef\refis##1{\indent\hbox to 0pt{\hss##1)~}}        
   \gdef\journal##1, ##2, ##3, ##4 {                     
      {\sl ##1~}{\bf ##2~}(##3) ##4 }}

 \def\refstyleprnp{              
   \gdef\refto##1{ (##1)}                                
   \gdef\refis##1{\indent\hbox to 0pt{\hss##1)~}}        
   \gdef\journal##1, ##2, ##3, 1##4##5##6{               
     {\sl ##1~}{\bf ##2~}(1##4##5##6) ##3}}

 \def\figurecaptions             
   {\endpage
    \beginparmode
    \head{Figure Captions}
 }

 \def\endpage                    
   {\vfill\eject}

 \def\endpaper                   
   {\endmode\vfill\supereject}


 \def\ref#1{Ref. #1}                     
 
 \def\frac#1#2{{\textstyle #1 \over \textstyle #2}}

 \def\sla{\raise.15ex\hbox{$/$}\kern-.57em}
 \def\leaderfill{\leaders\hbox to 1em{\hss.\hss}\hfill}
 \def\twiddle{\lower.9ex\rlap{$\kern-.1em\scriptstyle\sim$}}
 \def\bigtwiddle{\lower1.ex\rlap{$\sim$}}
 \def\gtwid{\mathrel{\raise.3ex\hbox{$>$\kern-.75em\lower1ex\hbox{$\sim$}}}}
 \def\ltwid{\mathrel{\raise.3ex\hbox{$<$\kern-.75em\lower1ex\hbox{$\sim$}}}}
 \def\square{\kern1pt\vbox{\hrule height 1.2pt\hbox{\vrule width 1.2pt\hskip
3pt
    \vbox{\vskip 6pt}\hskip 3pt\vrule width 0.6pt}\hrule height 0.6pt}\kern1pt}

\font\bigtenrm=cmbx10 scaled\magstep 2

\def\A{{\rm A}}
\def\e{\, {\rm e}}
\def\MT{{{\cal M}_3}}
\def\M4{{{\cal M}_4}}
\def\sqr#1#2{{\vcenter{\vbox{\hrule height.#2pt
        \hbox{\vrule width.#2pt height#1pt \kern#1pt
           \vrule width.#2pt}
        \hrule height.#2pt}}}}
\def\square{\mathchoice\sqr78\sqr78\sqr{6.1}7\sqr{5.5}7}
\def\IR{{\rm I}\!{\rm R}}
\def\inbar{\,\vrule height1.5ex width.4pt depth0pt}
\def\IC{\relax\hbox{$\inbar\kern-.3em{\rm C}$}}
\def\IZ{\relax\ifmmode\mathchoice
{\hbox{Z\kern-.4em Z}}{\hbox{Z\kern-.4em Z}}
{\lower.9pt\hbox{Z\kern-.4em Z}}
{\lower1.2pt\hbox{Z\kern-.4em Z}}\else{Z\kern-.4em Z}\fi}

\singlespace
\rightline{MITCTP\#2326}
\rightline{UBCTP-94-004}
\line{\hfill}
\rightline{July, 1994}
\vskip 0.4truein

{\let\bf=\bigtenrm
\doublespace
\centerline{\bf CANONICAL BF-TYPE TOPOLOGICAL FIELD THEORY}
\centerline{\bf AND FRACTIONAL STATISTICS OF STRINGS}}
\vskip 0.4truein

\centerline{{\bf Mario Bergeron}\footnote{*}{\tenpoint Work supported in
part by the Natural Sciences and Engineering Research Council of
Canada.}$^,$\footnote{**}{\tenpoint Work supported in
part by funds provided by the U.S. Department of Energy (D.E.O.) under
the contract \#DE-AC02-76ER03069 and the cooperative agreement
\#DE-FC02-94ER40818.}}
\vskip 0.2truein

\centerline{\it Center for Theoretical Physics, Laboratory for Nuclear
Science and Department of Physics}
\centerline{\it Massachusetts Institute of Technology,
Cambridge, Massachusetts, 02139 U. S. A.}
\vskip 0.3truein

\centerline{{\bf Gordon W. Semenoff} $^*$ and {\bf Richard J. Szabo}$^*$}
\vskip 0.2truein

\centerline{\it Department of Physics, University of British Columbia}
\centerline{\it Vancouver, British Columbia, V6T 1Z1 Canada}
\vskip 0.4truein

\centerline{Submitted to {\it Nuclear Physics} {\bf B}}
\vskip 0.4truein

\centerline{\bf ABSTRACT}
\vskip 0.1truein

We consider BF-type topological field theory coupled to non-dynamical
particle and string sources on spacetime manifolds of the form
$\IR^1\times\MT$, where $\MT$ is a 3-manifold without boundary. Canonical
quantization of the theory is carried out in the Hamiltonian formalism
and explicit solutions of the Schr\"odinger equation are obtained. We show
that the Hilbert space is finite dimensional and the physical states carry
a one-dimensional projective representation of the local gauge symmetries.
When $\MT$ is homologically non-trivial the wavefunctions in addition carry a
multi-dimensional projective representation, in terms of the linking matrix
of the homology cycles of $\MT$, of the discrete group of large gauge
transformations. The wavefunctions also carry a one-dimensional representation
of the non-trivial linking of the particle trajectories and string surfaces in
$\MT$. This topological field theory therefore provides a
phenomenological generalization of anyons to (3 + 1) dimensions where the
holonomies representing fractional statistics arise from the adiabatic
transport of particles around strings. We also discuss a duality between
large gauge transformations and these linking operations around the
homology cycles of $\MT$, and show that this canonical quantum field theory
provides novel quantum representations of the cohomology of $\MT$ and its
associated motion group.

\vfill\eject

\oneandahalfspace

\head{1. Introduction}

In (2 + 1) dimensional spacetime, anyons have a by now well-known physical
realization where magnetic flux tubes are attached to charged
particles and the Aharonov-Bohm phases resulting from adiabatic
transport of the composites give them fractional exchange statistics (see
[1] for a review). Anyons are also known to be phenomenologically described
in quantum field theory by Chern-Simons gauge theory, a topological field
theory [1,2]. We would expect these fractional phases to survive in a
4 dimensional world where flux tubes have infinite extent, as in the
conventional Aharanov-Bohm effect. In this Paper we shall study a
particular topological quantum field theory which gives
the appropriate generalization of anyons to (3 + 1) dimensions.

Holonomy effects in physical systems should always be described by some
topological field theory. In the following we shall consider a certain class of
Schwarz-type topological gauge theories, the so-called BF-theories [2--5],
which describe the type of holonomy which can occur in adiabatic transport
in a theory of point charges and strings where all other degrees of freedom
decouple. Abelian BF theories in 4 dimensions have for a long time been known
to be especially suited to describe physical systems where the sources are
vortex-like configurations, and it is well-known that there is a duality
between dynamical BF-type gauge theories and theories involving
Nambu-Goldstone fields with global abelian symmetries. This fact has been
used to describe an alternative approach to the Higgs mechanism, and it leads
to the London constitutive equation in superconductivity [6]. It has also
recently been exploited in dual models of cosmic strings [7], and axionic
black hole theories [8] where the axion charge is physically detectable
only by external cosmic strings in a 4 dimensional Aharanov-Bohm type process
[9] identical to what we shall consider here. These 4 dimensional
Aharanov-Bohm phases also appear in field theories of the QCD string [10].

On the mathematical side, it was realized in the late 1970's by Schwarz that
both abelian and nonabelian BF-theories in $d$ dimensions give effectively
computable path integral representations of a particular topological invariant
of $d$-manifolds, the Ray-Singer analytic torsion [2,3,5,11]. The properties
of this non-trivial topological invariant thus have natural interpretations
through standard techniques of quantum field theory. Furthermore, just as
Chern-Simons theory gives representations of linking numbers of curves in
3 dimensions [2], BF theories provide path integral representations of the
linking and intersection numbers of generic surfaces in $d$ dimensions
[2,3,5].

Not unrelated to these properties of BF theory is the fact that the statistical
phases which arise in some models of heterotic string compactification
illustrate explicitly the possibility of unusual statistics in 4 dimensional
string theory [12]. These statistical phases can be seen to arise from
certain cosmic string and superstring phenomena, such as anomaly
cancellation in 4 dimensions [13], and they arise from BF-type theories of
superstrings such as Nambu-Goto string theory modified with the inclusion
of the Kalb-Ramond term [14]. This possibility was made explicit by Aneziris
{\it et al.} who showed that more general statistics can exist for strings
in (3 + 1) dimensions [15]. Just as for particles moving on a plane where
the configuration space monodromies lead one to consider
the braid group of the plane, whose representations allow for exotic
statistics of charged particles in (2 + 1) dimensions [1], Aneziris {\it et
al.}
constructed representations of the holonomies of closed paths in the
configuration space of a system of identical, oriented strings which
demonstrate the possibility of fractional statistics.

The idea of fractional statistics for a system of identical strings
requires a better understanding of the Hilbert space of physical states for
such a system [12]. In this Paper we shall consider (3 + 1) dimensional abelian
BF-type topological field theory coupled to non-dynamical string sources, but
in the limiting case where one set of strings can be considered as points
in space (the situation in cosmic string scenarios) so that the composite
states are point particles with infinitely thin flux tubes attached (Section
2), the same situation that arises in pure Chern-Simons gauge theory. One
advantage of this simplifying limit is that we need not worry about
regulating the usual divergences which arise from self-linking terms.
We shall also consider only spacetime manifolds without boundary\footnote
{$^1$}{\tenpoint For spacetime manifolds with boundary, the analogs of
the boundary localized edge states and vertex operators of Chern-Simons
theory have been studied in [16].}.

We explicitly construct the Hilbert space of the theory by considering the
canonical quantization of this model in the Hamiltonian formalism. We begin
by examining how particle-string holonomies are represented in the
wavefunctions when the spatial 3-manifold is flat Euclidean 3-space $\IR^3$
(Section 3). There we obtain a simple representation of the particle-string
linking in $\IR^3$ through a spherical solid angle function which is the
3-dimensional version of the usual multivalued angle function which appears
in planar anyon theories [1]. In this case the Hilbert space is one dimensional
and the physical states carry a unitary representation of both the
statistics and the local gauge group.

We then consider the case where the spatial manifold is curved,
and also allow for non-trivial spatial topology (Section 4). There we obtain a
generalization of the adiabatic particle-string linking found in Euclidean
space through a generalized curved space spherical angle function, and the
wavefunctions again carry a one dimensional unitary representation of the
fractional statistics and of the local gauge symmetries. When the space has
non-trivial homology, we find that it is necessary to normalize the
cohomological parts of the gauge fields in terms of the linking matrix
of the homology cycles [17] in order for the correct particle-string
linking representation to appear. We shall also see that, when the coefficient
of the BF action (the ``statistics parameter") is of the form $Mk_1/k_2$ with
$k_1$ and $k_2$ integers and $M$ the integer-valued determinant  of the
linking matrix of the $p$ homology 1-cycles with the $p$ homology 2-cycles of
the 3-manifold, the Hilbert space is $|k_1k_2 M|^p$ dimensional and the
physical states in addition carry a $|k_2|^p$ dimensional projective
representation of the associated algebra of large gauge transformations.
This algebra and the ensuing representations are determined by the linking
matrix of the space, and thus the BF theory leads to a novel, complete
quantum representation of its homological properties. We shall further show
that
these large gauge transformations are dual to the linkings of particles
and strings around the non-trivial homology cycles of the manifold, and
their representation within the physical states provides a $|k_1|^p$
dimensional projective representation of the first and second cohomology
of the underlying space. The wavefunctions thus also yield intriguing
representations of the fundamental group of the particle-string
configuration space.

These results shed light on the structure of the enlarged Hilbert space for
4 dimensional strings induced by the more complicated fundamental group of
the string configuration space. They also illustrate the complexity of the
quantum theory of anyonic systems in (3 + 1) dimensions, and show as well
the rich mathematical structure inherent in BF theory insofar as providing
topological invariants in 3 dimensions. Some
details of the calculations in the case of a curved manifold are summarized
in an Appendix at the end of the Paper, where we derive an explicit
expression for the generalized adiabatic linking function.

\vfill\eject

\head{2. Canonical BF Theory in Four Dimensions}

We begin by describing some of the general features of 4-dimensional BF-type
topological field theory and its canonical formalism when coupled to sources.
Consider the field theory of a real-valued 2-form field $B$ and a
real-valued 1-form field $A$ defined on a 4 dimensional
spacetime manifold ${{\cal M}_4}$ with metric of Minkowski signature.
The BF action is given by the spacetime integral of a 4-form [2--5]
$$ S=\int_{{\cal M}_4}~{k\over2\pi}B\wedge dA \eqno(2.1)$$
where $1\over k$ is the coupling constant. This action is invariant under the
gauge transform
$$ A\rightarrow A+\chi\eqno(2.2) $$
where $\chi $ is a closed 1-form, $d\chi=0$. Modulo elements of the first
cohomology group of ${\cal M}_4$ this is trivially satisfied by an exact 1-form
$\chi=d\chi'$. On the other hand, like Chern-Simons theory in (2 + 1)
dimensions [2] the action transforms by a surface term under
$$ B\rightarrow B+\xi\eqno(2.3) $$
where $\xi$ is a closed 2-form, $d\xi=0$, which is also satisfied by an exact
2-form $\xi=d\xi'$ modulo elements of the second cohomology group of $\M4$.

In the present field theory without sources any closed forms are allowed in
(2.2) and (2.3). However, when this topological field theory is coupled to
sources we also require gauge invariance of the Wilson operators
$$W[L]=\exp\left(i\int_LA\right)\qquad,\qquad W[{\mit\Sigma}]
=\exp\left(i\int_{\mit\Sigma}B\right)\eqno(2.4)$$
for any oriented loop $L$ and any compact orientable
surface $\mit\Sigma$ in ${\cal M}_4$. This restricts the cohomological parts
of the closed forms allowed in (2.2) and (2.3) to those with integer-valued
cohomology so that
$$\int_L\chi=2\pi n\qquad,\qquad\int_{\mit\Sigma}\xi=2\pi m\eqno(2.5)$$
for some integers $n$ and $m$. In the following we shall
assume this restricted gauge symmetry.

The partition function is given by the path integral
$$Z=\int DA~DB~\exp\left(i\int_{{\cal M}_4}{k\over2\pi}B\wedge dA\right)
\eqno(2.6)$$
The functional integrals in (2.6) are performed over all gauge
orbits and are normalized by the volume of
the gauge group. This path integral is related to the
Ray-Singer analytic torsion [18] which is a topological invariant of
${{\cal M}_4}$ given by properties of the spectrum of the differential
operators $d$ and $\star d$ and the
Laplacian $\square_n$ acting on $n$-forms on $\M4$, and here is given
explicitly by the ratio of determinants [2,3,11]
$$Z={\det_{\perp}^{1/2}\square_1\over\det_{\perp}^{1/2}\square_0\det_\perp
^{1/4}\square_2}\eqno(2.7)$$
where $\det_\perp$ denotes the determinant with zero modes arising from
gauge invariance excluded.

Gauge- and topologically-invariant operators in this quantum field theory
are given by 2-cycle holonomies
of $B$ and 1-cycle holonomies of $A$. The expectation value of the Wilson loop
and surface variables (2.4) is given by the path integral with sources
$$ <W[L],W[{\mit\Sigma}]>~=~{\int DA\,DB\,
{}~\exp\left(i\int_{{\cal M}_4}~{k\over2\pi}B\wedge dA+i\int_L A+ i\int_
{\mit\Sigma}B\right)\over\int DA\,DB\,~\exp\left(i\int_{{\cal M}_4}~
{k\over2\pi}B\wedge dA\right)} \eqno(2.8)$$
where the Wilson loops are the world-lines
$L$ of particles and surfaces are the world-sheets $\mit\Sigma$ of
strings.  This integral is independent of the metric of ${{\cal M}_4}$
and is also formally a topological invariant.  It is related to the
topological linking number of surfaces $\mit\Sigma $ with contours $L$.
This can be seen by fixing the gauge and performing the integral to obtain
[2,3,5]
$$<W[L],W[{\mit\Sigma}]>~=\exp\left(-{2i\over \pi k}\int_
{\mit\Sigma}d{\mit\Sigma}_{\mu\nu}(x)~\int_L dl_\sigma (y)~
\epsilon^{\mu\nu\sigma\rho}~{ (x-y)_\rho\over\vert x-y\vert^4}\right)
=\e^{-{2\pi i \over k}I({\mit\Sigma},L)}\eqno(2.9)$$
where $d{\mit\Sigma}_{\mu\nu}(x)=d^2\sigma~\epsilon^{\alpha\beta}{\partial
x_\mu\over\partial\sigma^\alpha}{\partial x_\nu\over\partial\sigma^\beta}$
is the differential string area element and $l_\sigma$ parametrizes the
particle world-line. Here we have for simplicity assumed that ${\cal M}_4=
\IR^1\times\IR^3$ and we use the convention $\epsilon^{0123}=+1$.
(2.9) yields the standard expression for the Gaussian linking number
$I({\mit\Sigma},L)$ of a contour $L$ and a surface ${\mit\Sigma}$ [19].
Path integral representations of these linking numbers in arbitrary
4-manifolds can also be obtained [2,3,5] (see the Appendix, equation (A.23)).

A more complete picture of this particle-string system is obtained by canonical
quantization.  For this we choose the spacetime to be the product
manifold $\IR^1\times{{\cal M}_3}$ where $\IR^1$ parametrizes the time.
The action is now\footnote{$^2$}{\tenpoint In this Paper we
shall implicitly assume that all antisymmetric tensors, Dirac delta
functions, spatial index contractions and volume forms $d^3x$ contain the
appropriate metric factors required for diffeomorphism invariance. Also, in the
following $*$ denotes the Hodge duality operator defined with respect to the
metric of $\MT$.}
$$ S=\int dt\,\int_{{\cal M}_3}d^3x\,
\left({k\over4\pi}\epsilon^{\mu\nu\rho\sigma} B_{\mu\nu}\partial_\rho
A_\sigma +A_\mu j^\mu+{1\over2}B_{\mu\nu}\Sigma^{\mu\nu}\right)\eqno(2.10) $$
where the particle current is given by
$$ j^\mu(x)=\sum_aq_a\int_{L_a}d\tau\,{dr_a^\mu(\tau)\over d\tau}
\delta^{(4)}(x,r_a(\tau))\eqno(2.11)$$
with $r_a^\mu(\tau)$ the embedding of the world line $L_a$ of particle
$a$ with charge $q_a$ in ${\cal M}_4$. It satisfies the continuity equation
$\partial_\mu j^\mu=0$ when the world lines are closed.
The string current is given by
$$\Sigma^{\mu\nu}(x)=\sum_b\phi_b\int_{{\mit\Sigma}
_b}d^2\sigma\,\epsilon^{\alpha\beta}
{\partial X_b^\mu(\sigma)\over\partial\sigma^\alpha}
{\partial X_b^\nu(\sigma)\over\partial\sigma^\beta}
\delta^{(4)}(x,X_b(\sigma))\eqno(2.12)$$
with $X_b^\mu(\sigma)$ the embedding of the world-sheet
${\mit\Sigma}_b$ of string $b$ with ``electromagnetic" flux $\phi_b$ in ${\cal
M}_4$.  It is antisymmetric, $\Sigma^{\mu\nu}=-\Sigma^{\nu\mu}$, and
conserved, $\partial_\mu\Sigma^{\mu\nu}=0$, when either the string
world-sheet is closed or the embedding obeys the appropriate boundary
conditions.

When the 3-manifold $\MT$ is compact, we require that these sources further
obey the property that the total charge and total flux both vanish,
$\sum_aq_a=\sum_b\phi_b=0$. The gauge
constraints (see (2.13) below) imply that the inclusion of
non-zero charge and flux sectors of the theory require the gauge connections
$A$ and $B$ to be sections of some vector bundles over $\MT$ (as then they
have non-vanishing curvature), and they can only be defined locally on patches
over $\MT$. In this case the BF action (2.1) must be modified appropriately
[20], and we shall not consider this technical adjustment
in this Paper. Intuitively, the constraint that the total charge and flux
vanish in a closed space is required since their electric and
``magnetic" fluxes have nowhere to go.

The temporal components of the fields are Lagrange
multipliers which enforce the constraints
$$ {k\over2\pi}\partial_iB^i+j^0\sim 0 ~~,~~ {k\over2\pi}
\epsilon^{ijk}\partial_jA_k+\Sigma^{0i}\sim 0 \eqno(2.13)$$
where $B^i(x)={1\over2}\epsilon^{ijk}B_{jk}(x)$. The second constraint in
(2.13) confines ``electromagnetic" flux to the string world sheets and gives
the
analog of the Meissner effect in a BCS superconductor [6]. The first constraint
then couples this flux to the particle charges with coupling $1\over k^2$.
Thus the propagating gauge degrees of freedom in this field theory decouple
and their only roles are to attach infinitely long flux tubes to the point
charges in this theory in terms of the coefficient $k$. As in
Chern-Simons theory [1], it is this coupling which leads to Aharanov-Bohm
phases
which are ultimately responsible for the statistics of these charged
particle composites.

The remaining action
$$ S=\int dt\,\int_{{\cal M}_3}d^3x\,\left( {k\over2\pi}B^i\dot A_i+A_i
j^i+{1\over2}B_{ij}\Sigma^{ij}\right)\eqno(2.14) $$
is of first order in time derivatives and is therefore already expressed in
phase space with the spatial components of $A$ and $B$ being the canonically
conjugate variables. It leads to the canonical commutator
$$ \bigl[ A_i(x),B^j(y)\bigr]={2\pi i\over k}\delta_i^j\delta^{(3)}(x,y)
\eqno(2.15)$$
In the temporal gauge $A_0=0$, $B_{0i}=0$, the Hamiltonian is
$$ H=\int_{{\cal
M}_3}d^3x\,\left(-A_ij^i-{1\over2}B_{ij}\Sigma^{ij}\right)\eqno(2.16)$$
The canonical formalism above can now be used to solve for the spectrum of the
Hamiltonian (2.16). In the following we shall show that the associated Hilbert
space states represent fractional particle-string linking numbers as well as
the possible non-trivial topological properties of the 3-manifold ${\cal M}_3$.

\head{3. Particle-String Holonomy in Euclidean Space}

To get an intuitive idea of how the wavefunctions of BF-theory represent
fractional linking numbers of particles and strings, we first consider the
relatively simple case where $\MT$ is Euclidean 3-space $\IR^3$.
We can decompose the fields over $\IR^3$ using the Hodge decomposition,
assuming that they have compact support. We then have the usual
longitudinal and transverse decompositions
$$A=d\theta+*dK^\prime~~{\rm
or}~~A_i=\nabla_i\theta+\epsilon_{ijk}\nabla^j {K^\prime}^k\eqno(3.1)$$
$$*B=d\theta^\prime+*dK~~{\rm or}~~B^i=\nabla^i\theta^\prime+\epsilon^{ijk}
\nabla_j K_k\eqno(3.2)$$
The commutation relations (2.15) now take the form
$$ \bigl[ \theta(x),\theta^\prime(y)\bigr]=-{2\pi i\over
k}{1\over \nabla^2}\delta^{(3)}(x,y)~~~,
{}~~~\bigl[ K_i(x),{K^\prime}^j(y)\bigr]={2\pi i\over k}\left(\delta_i^j-
{\nabla_i\nabla^j\over\nabla^2}\right){1\over\nabla^2}\delta^{(3)}(x,y)
\eqno(3.3)$$
and can be represented by the functional derivative operators
$$\eqalign{A_i(x)&=\nabla_i\theta(x)-{2\pi i\over k}\epsilon_{ijk}\nabla^j
{1\over \nabla^2}{\delta \over \delta K_k(x)}\cr
B^i(x)&=\epsilon^{ijk}\nabla_j K_k(x)+{2\pi i\over k}\nabla^i
{1\over\nabla^2}{\delta\over\delta\theta(x)}\cr}\eqno(3.4)$$

We treat the constraints (2.13) as physical state conditions which
separate the gauge-invariant states in the Hilbert space of the
canonical quantum field theory. In the functional Schr\"odinger picture, we
see from (3.4) that they are solved by wavefunctionals of the form
$$\Psi_{\rm phys}[\theta,K;t]=\exp\left[i\int d^3x\,\left(
\theta(x)j^0(x,t)+K_i(x)\Sigma^{0i}(x,t)\right)\right]\tilde\Psi(t)\eqno(3.5)$$
Note that the continuity equation $\partial_\mu\Sigma^{\mu
0}=0$ is required for the constraints (2.13) to be satisfied in (3.5) here.
The time-independent gauge symmetries (2.2) and (2.3) are therefore
represented projectively as
$$\Psi_{\rm phys}[\theta+\chi',K+\xi';t]=\exp\left[i\int d^3x~\left(\chi'(x)
j^0(x,t)+\xi_i'(x)\Sigma^{0i}(x,t)\right)\right]\Psi_{\rm phys}[
\theta,K;t]\eqno(3.6)$$
where the projective phase is a non-trivial local ${\rm U}(1)$ 1-cocycle.
Then, to solve the Schr\"odinger equation $$ i{\partial\over\partial
t}\Psi_{\rm phys}[\theta,K;t]= H\Psi_{\rm phys}[\theta,K;t]\eqno(3.7)$$
from (2.16) and (3.4) we obtain
$$\tilde{\Psi}(t)=\exp\biggl[{2\pi i\over k}\int_{-\infty}^t dt'\,\int d^3x\,
\biggl((\nabla\times j)_i{1\over\nabla^2}\Sigma^{0i}(x,t')
+(\nabla\cdot\Sigma){1\over\nabla^2}j^0(x,t')\biggr)\biggr]\eqno(3.8)$$
where $\Sigma_i={1\over2}\epsilon_{ijk}\Sigma^{jk}$.

We now substitute in the source currents (2.11) and (2.12) with the particle
trajectories and string surfaces in $\MT$ parametrized by time; i.e.
$r_a^0(\tau)=\tau$ and $X_b^0(\sigma^1,\sigma^2)=\sigma^1$. Then
using the Green's function for the 3 dimensional Laplacian on $\IR^3$
$$(x|{1\over\nabla^2}|y)=-{1\over4\pi|x-y|}\eqno(3.9)$$
and integrating by parts over $\IR^3$ the full wavefunctional is therefore
$$\eqalign{\Psi_{\rm phys}[\theta,K;t]&=\exp\biggl[i
\left\{\sum_aq_a\theta(r_a(t))+\sum_b\phi_b\int d\sigma\,{\partial
X_b^i(t,\sigma)\over\partial\sigma}K_i(X_b(t,\sigma))\right\}\cr&\qquad
+{i\over2k}\sum_{a,b}q_a\phi_b\int_{-\infty}^tdt'\,
\int d\sigma\,\epsilon_{ijk}\biggl\{\dot{r}_a^i(t'){\partial X_b^k(t',\sigma)
\over\partial\sigma}\cr&\qquad-{\partial X_b^i(t',\sigma)\over\partial t'}
{\partial X_b^k(t',\sigma)\over\partial\sigma}
\biggr\}{\left(r_a(t')-X_b(t',\sigma)\right)^j\over|r_a(t')
-X_b(t',\sigma)|^3}\biggr]\cr}\eqno(3.10)$$

The final sum in (3.10) (i.e. $\tilde{\Psi}(t)$) is the topological
linking number of the particle trajectories and the string
surfaces in $\IR^3$ and is the three dimensional
version of the linking number in (2.9). This canonical linking is the adiabatic
limit of the covariant linking found in (2.9) using the path integral approach.
As a function of $t$ it is the analog of the usual 2 dimensional multivalued
angle function which arises in anyon theories as the term responsible for
giving charged particles anomalous exchange statistics [1].
If we look closely at the integrand of this term, we see that it has
the form
$${R_{ab}\cdot(dR_{ab}\times dR_{ab})
\over|R_{ab}|^3}=d\Phi_{ab}(t)\eqno(3.11)$$
where $R_{ab}=X_b-r_a$ and $\Phi_{ab}(t)$ is the solid
angle formed by the string (along $X_b(t,\sigma)$)
as viewed from the charged particle (at $r_a(t)$). Here the ``statistics
parameter" is $k\over2\pi$, i.e. when a particle of charge $q_a$
encircles a string of flux $\phi_b$ once, $\Psi_{\rm phys}$ changes by
the phase $\e^{-{2\pi i\over k}q_a\phi_b}$. The Hilbert space of BF-theory on
Euclidean 3-space is therefore 1 dimensional and the wavefunctionals carry
a 1 dimensional unitary representation of the fractional particle-string
linking for generic values of the BF coefficient $k$.

\head{4. Particle-String Holonomy in Curved Spaces}

We now move on to the more interesting case of a curved space with
possibly non-trivial topology. We assume that $\MT$ is a compact,
path-connected, orientable 3-manifold without boundary.
The non-trivial homology of $\MT$ is represented by the homology groups
$H_1(\MT;\IZ)$ and $H_2(\MT;\IZ)$, which by Poincar\'e-Hodge duality both
have the same dimension that we denote by $p$. We shall ignore the torsion
parts of the homology groups which play no role in the following. Let
$\{L^m\}_{m=1}^p$ be a set of generators of $H_1(\MT;\IZ)$ and $\{{\mit\Sigma}
^m\}_{m=1}^p$ a set of generators of $H_2(\MT;\IZ)$. A fundamental
characteristic of $\MT$ is its linking matrix [17]
$$M^{mn}=I_{\MT}({\mit\Sigma}^m,L^n)=\sum_{I_{mn}}{\rm
sgn}~(I_{mn})\eqno(4.1)$$
which counts the signed intersections $I_{mn}$ of the two dimensional surface
${\mit\Sigma}^m$ with the one dimensional line $L^n$ in $\MT$. By definition
it is a topological invariant and an integer-valued matrix. Note that it
is not necessarily symmetric and so it is important to keep track of the
order of its two indices in what follows. Some examples of constant curvature
spaces are the 3-sphere $S^3$ for which $p=0$, the 3-torus $(S^1)^3$ where
$p=3$, and the product 3-manifold $S^2\times S^1$ wherein $p=1$. Notice that
the linking matrix $M^{nm}$ for the latter 2 examples is just the $p\times p$
identity matrix, as is always the case for a space $\MT$ which is the product
of a one- and a two-dimensional space.

To describe the non-trivial cohomology of $\MT$, we
let $\{\alpha_\ell\}_{\ell=1}^p\in H^1(\MT;\IR)$ be a basis
of harmonic 1-forms (i.e. $d\alpha_\ell=d*\alpha_\ell=0$) and $\{\beta_\ell\}_
{\ell=1}^p\in H^2(\MT;\IR)$ a basis of harmonic 2-forms which are
Poincar\'e-duals of the homology generators
$$
\int_{L^m}\alpha_\ell=\delta_\ell^m~~~~,~~~~\int_{{\mit\Sigma}^m}\beta_\ell
=\delta_\ell^m\eqno(4.2)
$$
As will become clear later on (see the Appendix, equation (A.8)), for our
purposes it is necessary to choose the normalization (4.2) instead of the
standard normalization usually employed wherein the harmonic generators
$\alpha_\ell$ and $\beta_\ell$ are taken to be Hodge-dual basis elements [2].
Here the exterior product of these cohomology generators satisfy the
reciprocal relation of the homology generators
$$\int_\MT\alpha_\ell\wedge\beta_k=M_{\ell k}\eqno(4.3)$$
where we define the inverse and determinant of the linking matrix as
$$
M_{\ell k}M^{k m}=\delta_\ell^m~~~~,~~~~M=\det[M^{mn}]\eqno(4.4)
$$
In general $M_{mn}$ is rational-valued, but $M\cdot M_{mn}$ is always
an integer-valued matrix. We will assume that $M$ is positive, since this
can always be achieved by changing the orientation of $\MT$ if
necessary. Finally, the non-Euclidean
geometry of $\MT$ is encoded within the Hodge duality operator $*$.

The Hodge decompositions of the forms $A$ and $B$ on $\MT$ are now as follows:
The 1-form $A=A_idx^i$ restricted to $\MT$ can be decomposed into exact,
co-exact and harmonic forms as
$$A=d\theta+*d K^\prime+a^\ell\alpha_\ell\eqno(4.5)$$
where the scalar field $\theta$ and 1-form field $K^\prime$ satisfy
$$\nabla_g^2\theta=*d*A~~,~~\Phi^\prime=*d*d K^\prime=*dA\eqno(4.6)$$
and the harmonic coefficients are given by
$$a^\ell(t)=M^{m\ell}\int_\MT A\wedge\beta_m=\int_{L^\ell}A\eqno(4.7)$$
Here $\nabla_g$ denotes the covariant derivative with respect to
the metric $g$ of $\MT$ and $\nabla_g^2$ is the corresponding scalar
Laplacian. Likewise the 2-form $B={1\over2}B_{ij}dx^i\wedge dx^j$ can be
decomposed on $\MT$ as
$$B=d K+*d\theta^\prime+b^\ell\beta_\ell\eqno(4.8)$$
where the scalar field $\theta^\prime$ and 1-form field $K$ obey
$$\nabla_g^2\theta^\prime=*dB~~,~~d*d K=d*B\eqno(4.9)$$
and the harmonic coefficients are
$$b^\ell(t)=M^{\ell m}\int_\MT B\wedge\alpha_m=\int_{{\mit\Sigma}
^\ell}B\eqno(4.10)$$

Since by assumption $\int_\MT d^3x~\nabla_g\cdot B=0$, the function $*dB$ in
(4.9) contains no zero modes, or the constant function in this case, of the
scalar Laplacian $\nabla_g^2$. Similarily the 1-form $*dA$ in (4.6)
contains no zero modes of the Laplacian $(\nabla_g^{(1)})^2$ acting on
1-forms. Moreover, we can set the zero modes of $d*B=\nabla_g\times B$ and
$*d*A=\nabla_g\cdot A$ to zero using time-independent gauge transformations.
The harmonic parts (4.7) and (4.10) of the fields represent the additional
degrees of freedom of $A$ and $B$ that are present when $\MT$ is
homologically non-trivial (compare with equations (3.1) and (3.2) of
Section 3).

The gauge constraints (2.13) are again solved by the decomposition (3.5) of the
wavefunctions. Substituting (4.5)--(4.10) into the remaining action, without
the gauge constraints, in (2.14) and integrating by parts over $\MT$ gives
$$S=\int dt\,\left[\int_\MT\left({k\over2\pi}(-\dot{\theta}
\nabla_g^2\theta^\prime+\dot{\Phi}^\prime_i K^i)~d^3x+A\wedge*
\tilde{j}+B\wedge*\tilde{\Sigma}\right)+{k\over2\pi}
\dot{a}^\ell M_{\ell m}b^m\right]\eqno(4.11)$$
where $\tilde{j}$ and $\tilde{\Sigma}$ are the dual forms, over
$\MT$, of the particle and string current vector fields (2.11) and
(2.12), respectively. From (4.11) we read off the non-vanishing canonical
commutators
$$\left[\theta(x),\nabla_g^2\theta^\prime(y)\right]=-{2\pi i\over k}
\delta^{(3)}(x,y)\quad,\quad\left[K_i(x),\Phi^\prime_j(y)\right]=
-{2\pi i\over k}P_{ij}\delta^{(3)}(x,y)\eqno(4.12)$$
$$\left[a^\ell,b^m\right]={2\pi i\over k}M^{\ell m}\eqno(4.13)$$
where $P_{ij}$ is the transverse projection operator satisfying
$$\nabla_g^iP_{ij}=\nabla_g^jP_{ij}=0~~~,~~~P_{ij}A^j=A_i-(\nabla_g)_i
\left({1\over(\nabla_g^2)^\prime}\nabla_g^jA_j\right)\eqno(4.14)$$
and the prime on the Laplacian indicates that we remove its zero modes.

In the Schr\"odinger picture we therefore consider
the wavefunctions $\Psi_{\rm phys}[\theta,K,a;t]$ and we can represent
the commutation relations (4.12) and (4.13) by the derivative operators
$$\theta^\prime(x)={2\pi i\over k}{1\over(\nabla_g^2)^\prime}{\delta\over
\delta\theta(x)}\quad,\quad\Phi_i^\prime(x)={2\pi i\over
k}P_{ij}{\delta\over\delta K_j(x)}\eqno(4.15)$$
$$b^m=-{2\pi i\over k}M^{\ell m}{\partial\over\partial a^\ell}\eqno(4.16)$$
As we shall see, the projection operator $P_{ij}$ ensures the invariance of
the wavefunctions under the time-independent longitudinal symmetry $K\to K+d
\Lambda$, where $\Lambda $ is an arbitrary function, which is dual to the
string current symmetry $\tilde{\Sigma}\to\tilde{\Sigma}+*d\tilde{\Lambda}$,
where $\tilde{\Lambda}$ is an arbitrary 1-form. This symmetry arises from the
continuity equation for $\Sigma^{\mu\nu}$.

\subhead{4.1. Source-free Case: Effects of Non-trivial Topology and
Cohomology Representations}

When $\MT$ has non-trivial homology, the coefficients (4.7) and (4.10)
of the harmonic parts of the gauge fields transform under the large gauge
transformations which arise from windings around the non-trivial homology
cycles of $\MT$ (i.e. $n,m\neq0$ in (2.5)). The physical state wavefunctions
should respect these global gauge symmetries just as they respect the local
gauge symmetries generated by (2.13) (see (3.6)), and this constraint then
partitions the physical Hilbert space into superselection sectors labelled
by the homology classes of $\MT$. To see how this works, we first consider
this model without sources. Then the constraints (2.13) imply that $A$ and
$B$ restricted to $\MT$ are closed forms. Consequently, the harmonic
coefficients (4.7) and (4.10) are topological invariants which generate the
signed intersection number (4.1) through the operator algebra
(4.13)\footnote{$^3$}{\tenpoint Had we instead chosen the normalization of
the harmonic generators corresponding to Hodge duality, i.e. $\int_\MT\alpha_
\ell\wedge\beta_k=\delta_{\ell k}$, then the operator algebra of $a^\ell$ and
$b^m$ would involve the identity matrix $\delta^{\ell m}$ rather than the
linking matrix $M^{\ell m}$. However, because of the canonical commutator
(2.15), the contour and surface integrals of $A$ and $B$, respectively, would
still satisfy the algebra (4.13).}.

With no sources present the Hamiltonian (2.16) vanishes and so the
wavefunctions
are time-independent. Moreover, the local gauge constraints (2.13) and
the relations (4.5), (4.8), and (4.15) imply that
$${\delta\over\delta\theta}\Psi_{\rm phys}[\theta,K,a]=P_{ij}
{\delta\over\delta K_j}\Psi_{\rm phys}[\theta,K,a]=0\eqno(4.17)$$
so that the physical states depend only on the global harmonic
coefficients $a^\ell$. It remains to solve for the invariance under the
time-independent large gauge transformations (2.2) and (2.3) satisfying (2.5),
which in terms of the harmonic coefficients (4.7) and (4.10) are given by the
translations
$$a^\ell\to a^\ell+2\pi n^\ell\quad,\quad b^\ell\to b^\ell+
2\pi m^\ell\quad;\quad n^\ell,m^\ell\in\IZ\eqno(4.18)$$

It is convenient here to instead work in a holomorphic polarization defined
by the complex variables\footnote{$^4$}{\tenpoint This polarization eludes the
delta-function singularities which appear in the wavefunctions when the usual
Schr\"odinger polarization is used.}
$$\gamma^\ell=a^\ell+M^{m\ell}\rho_{mk}b^k\quad,\quad
\overline{\gamma}^\ell=a^\ell+M^{m\ell}\overline{\rho}_{mk}b^k\eqno(4.19)$$
where $\rho=[\rho_{mk}]$ is an arbitrary symmetric $p\times p$
complex-valued matrix whose imaginary part is negative-definite. The gauge
transformations (4.18) in these new variables are
$$\gamma^\ell\to\gamma^\ell+2\pi(n^\ell+M^{m\ell}\rho_{mk}m^k)\quad,\quad
\overline{\gamma}^\ell\to\overline{\gamma}^\ell+2\pi(n^\ell
+M^{m\ell}\overline{\rho}_{mk}m^k)\eqno(4.20)$$
and the commutation relations (4.13) become
$$\left[\gamma^\ell,\gamma^k\right]=\left[\overline{\gamma}^\ell,
\overline{\gamma}^k\right]=0\qquad,\qquad\left[\gamma^\ell,
\overline{\gamma}^k\right]=-{2\pi\over k}\Omega^{\ell k}\eqno(4.21)$$
where
$$\Omega^{\ell k}=-2\,M^{p\ell}~{\rm Im}\,\rho_{pq}\,M^{qk}\eqno(4.22)$$
is a real-valued positive-definite symmetric matrix. These relations may be
represented by $$\overline{\gamma}^\ell={2\pi\over k}\Omega^{\ell k}
{\partial\over\partial\gamma^k}\eqno(4.23)$$
and in the coordinates (4.19) we will instead be looking for the coherent
state wavefunctions which we denote by $\Psi_0(\gamma)$.

The matrix $\rho$ introduced above can be interpreted as follows.
The phase space of the source-free BF theory is the $2p$ dimensional
space of the $a^\ell$ and $b^\ell$ variables
$${\cal P}=H^1(\MT;\IR)\oplus H^2(\MT;\IR)\eqno(4.24)$$
$\rho$ can then be thought of as parametrizing a complex structure
on ${\cal P}$ forming the $p$ dimensional complex space of the
$\gamma^\ell$ variables, and this then determines all of the topological
degrees
of freedom which remain in the source-free case modulo the large gauge
transformations (4.20). In fact, as we shall see later on, the
positive-definite symmetric matrix $\Omega$ actually defines a
metric on $\cal P$, and from its definition (4.22) (and the definition (4.19))
we see that it incorporates the topological linking of the homology cycles of
$\MT$. However, since the action (2.1) defines a topological
field theory, all observables will be independent of this phase space
complex structure. This is analogous to the situation in Chern-Simons theory
[21]. From (4.24) and (4.19) we see that the quantization of the classical
phase space of the BF system will give (projective) quantum representations
of the cohomology groups of $\MT$.

{}From (4.21) it follows that the quantum operators which generate the
global gauge transformations (4.20) in the Schr\"odinger picture are
$$U(n,m)=\exp\left[2\pi\left(n^\ell+M^{m\ell}\rho_{mk}m^k\right)
{\partial\over\partial\gamma^\ell}-k(n^\ell+M^{m\ell}\overline{\rho}_{mk}m^k)
(\Omega^{-1})_{\ell q}\gamma^q\right]\eqno(4.25)$$
The unitary operators (4.25) in general do not commute among themselves,
in contrast with their classical counterparts, and the
Baker-Campbell-Hausdorff formula
$$\e^{X+Y}=\e^{-[X,Y]/2}\e^X\e^Y\eqno(4.26)$$
along with the commutation relations (4.21) show that the algebra of these
large gauge transformations is a variation of a clock algebra
$$U(n,m)U(n',m')=\e^{2\pi
ik(n'^\ell M_{\ell n}m^n-n^\ell M_{\ell n}m'^n)}U(n',m')U(n,m)\eqno(4.27)$$
This ${\rm U}(1)$ 2-cocycle relation differs from the standard clock algebra
in that the matrix $M_{\ell n}$ in (4.27) is usually the identity matrix
$\delta_{\ell n}$. On 3-manifolds $\MT$ for which the linking matrix is not
trivially the identity matrix the 2-cocycle appearing in (4.27) is quite
natural, since it then also reflects the possible non-trivial linkings of
the homology cycles of $\MT$ which should be represented by the generators of
windings around them. To determine the effect of these generators explicitly
on the wavefunctions $\Psi_0(\gamma)$, we separate out the derivative
part of the operator (4.25) using the Baker-Campbell-Hausdorff formula (4.26)
again, and we find that the behaviour of the wavefunctions under large gauge
transformations is
$$\eqalign{U(n,m)\Psi_0(\gamma^\ell)=&\exp\Bigr[-k(n^\ell+M^{m\ell}
\overline{\rho}_{mk}m^k)(\Omega^{-1})_{\ell q}\gamma^q-\pi
k(n^\ell+M^{m\ell}\overline{\rho}_{mk}m^k)(\Omega^{-1})_{\ell q}\cr&
\qquad\qquad\times(n^q+M^{rq}\rho_{rs}m^s)\Bigl]\Psi_0\left(\gamma^\ell+2
\pi(n^\ell+M^{m\ell}\rho_{mk}m^k)\right)\cr}\eqno(4.28)$$

When the coupling constant $1\over k$ is an irrational number, the algebra
(4.27) of large gauge transformations is infinite dimensional. Finite
dimensional representations of the non-trivial homology of $\MT$ can
however be obtained when the BF coefficent $k$ is either
a positive integer or rational number, $k=M{k_1\over k_2}$; $k_1,\,k_2\in
\IZ^+$ (Negative $k$ can then be obtained by reversing the orientation of
$\MT$ and thus changing the sign of $M$). We shall assume these
discrete values of $k$ for the remainder of this Paper. Then for any set of
integers $(n,m)$, the algebra (4.27) implies that $U(k_2n,k_2m)$ commutes with
all of the other gauge transformation generators, as then the phase in (4.27)
is an integer multiple of $2\pi$. Since these operators are unitary
and commute with the (zero) Hamiltonian here, it follows that their
action (4.28) on the wavefunctions must lie on the same ray in the
Hilbert space as that defined by $\Psi_0(\gamma)$, i.e.
$$U(k_2n,k_2m)\Psi_0(\gamma)=\e^{i\eta_{(n,m)}}\Psi_0(\gamma)\eqno(4.29)$$
for some phases $\eta_{(n,m)}\in S^1$. Comparing this with (4.28) we see
that the wavefunctions enjoy the quasi-periodicity property
$$\eqalign{\Psi_0\bigl(\gamma^\ell+2\pi
k_2(n^\ell+M^{m\ell}&\rho_{mk}m^k)\bigr)
=\exp\Bigl[i\eta_{(n,m)}+k_1(n^\ell+M^{m\ell}\overline{\rho}_{mk}m^k)
(\Omega^{-1})_{\ell q}\gamma^q\cr&+\pi k_1k_2(n^\ell+M^{m\ell}\overline{\rho}_
{mk}m^k)(\Omega^{-1})_{\ell q}(n^q+M^{rq}\rho_{rs}m^s)\Bigr]\Psi_0(\gamma^
\ell)\cr}\eqno(4.30)$$

The only functions which obey quasi-periodic conditions such as
(4.30) are combinations of the Jacobi theta functions [22]
$$\Theta\pmatrix{c\cr d\cr}(z|\Pi)=\sum_{\{n_\ell\}\in\IZ^p}
\exp\left[i\pi(n^\ell+c^\ell)\Pi_{\ell k}(n^k+c^k)+2\pi i
(n^\ell+c^\ell)(z_\ell+d_\ell)\right]\eqno(4.31)$$
where $c^\ell,\,d_\ell\in[0,1]$ and $z_\ell\in\IC$.
The functions (4.31) are well-defined holomorphic functions of
$\{z_\ell\}\in\IC^p$ for $\Pi=[\Pi_{\ell k}]$ in the
Siegal upper half-plane, and they obey the doubly semi-periodic conditions
$$\Theta\pmatrix{c\cr d\cr}(z_\ell+s_\ell+\Pi_{\ell k}t^k|\Pi)=
\exp\left[2\pi ic^\ell s_\ell-i\pi t^\ell\Pi_{\ell k}t^k-2\pi it^\ell(z_\ell
+d_\ell)\right]\Theta\pmatrix{c\cr d\cr}(z|\Pi)\eqno(4.32)$$
where $s_\ell$ and $t^\ell$ are integers, and
$$\Theta\pmatrix{c\cr d\cr}(z_\ell+C\Pi_{\ell k}t^k|\Pi)=\exp\left[-i\pi C^2
t^\ell\Pi_{\ell k}t^k-2\pi iCt^\ell(z_\ell+d_\ell)\right]\Theta\pmatrix
{c+Ct\cr d\cr}(z|\Pi)\eqno(4.33)$$
for any non-integer $C\in\IR$. It should be emphasized that the
transformations (4.32) can be performed in many different steps with the
same final result, but successive applications of (4.32) and
(4.33) do not commute. Instead, when applied in different orders, the
final results differ by a phase which forms a representation of the algebra
(4.27). We {\it define} the operators $U(n,m)$ with the convention that the
transformation (4.32) is applied before (4.33).

After some algebra, we find that the algebraic constraints (4.30)
are uniquely solved by the wavefunctions
$$\Psi_0^{(q)}\pmatrix{c\cr d\cr}(\gamma)=\e^{{k\over4\pi}
\gamma^\ell(\Omega^{-1})_{\ell k}\gamma^k}\Theta\pmatrix{{c+
q\over Mk_1k_2}\cr d\cr}\left({Mk_1\over2\pi}M_{m\ell}\gamma^m\Bigm|
-k_1k_2M\rho\right)\eqno(4.34)$$
where $q^\ell=1,\,2,\dots,k_1k_2M$. Computing the periodic behaviour of (4.34)
we find that the phases in (4.29) are given by the non-trivial 1-cocycle
$${\eta_{(n,m)}\over2\pi}=\alpha_1(n,m)=c^\ell M_{m\ell}n^m+d_\ell
m^\ell-{1\over2} k_1k_2Mn^mM_{m\ell} m^\ell\eqno(4.35)$$
Under an arbitrary large gauge transformation (4.28) we have
$$\eqalign{U(n,m)\Psi_0^{(q)}\pmatrix{c\cr d\cr}(\gamma)&=\exp\left[
{2\pi i\over k_2}\left(c^\ell M_{m\ell}n^m+d_\ell m^\ell\right)-i\pi
kn^mM_{m\ell}m^\ell+{2\pi i\over k_2}q^\ell M_{m\ell}
n^m\right]\cr&\qquad\qquad\qquad\qquad\times\Psi_0^{(q^\ell-k_1 M m^\ell)}
\pmatrix{c\cr d\cr}(\gamma)\cr&=\sum_{q'}[U(n,m)]_{qq'}\Psi_0^{(q')}
\pmatrix{c\cr d\cr}(\gamma)\cr}\eqno(4.36)$$
where the unitary matrices
$$[U(n,m)]_{qq'}=\exp\left[{2\pi i\over k_2}\left(c^\ell M_{m\ell}n^m+
d_\ell m^\ell\right)-i\pi kn^mM_{m\ell} m^\ell+{2\pi i\over k_2}q^\ell
M_{m\ell}n^m\right]\delta_{q^\ell-k_1 M m^\ell,q'^k}\eqno(4.37)$$
form a $(k_2)^p$ dimensional projective representation, which is cyclic of
period $k_2$, of the algebra (4.27) of large gauge transformations. The
projective phase here is the non-trivial global ${\rm U}(1)$ 1-cocycle
$$\alpha_1^{(q)}(n,m)={1\over k_2}\left(c^\ell M_{r\ell}n^r+d_\ell m^\ell
+q^\ell M_{r\ell}n^r-{Mk_1\over2}n^rM_{r\ell}m^\ell\right)\eqno(4.38)$$

Therefore, the Hilbert space of source-free BF-theory,
on spaces with $p$ dimensional first and second homology and linking matrix
$M^{k\ell}$, is $(k_1k_2M)^p$ dimensional and the wavefunctions carry a
$(k_2)^p$ dimensional representation of the corresponding discrete group
of large gauge transformations. Notice that the possible non-trivial
homology linking is even represented directly in the dimensionality of the
Hilbert space.

\subhead{4.2. Inclusion of Sources: Fractional Statistics and Duality}

In the last Subsection we have seen that on homologically non-trivial spaces,
the Hilbert space of physical states is non-trivial even in the absence of
particle and string sources (in contrast to the simply connected case), due to
the existence of non-trivial global gauge transformations.
To obtain representations of the particle-string holonomy in $\MT$, we
re-introduce sources into the problem and Hodge decompose the
dual current forms $\tilde{j}=j_idx^i$ and $\tilde{\Sigma}=
{1\over2}\Sigma_{ij}dx^i\wedge dx^j$ as before on $\MT$. With the same
conventions for the particle trajectories and string surfaces as in
Section 3 above, we have
$$\tilde{j}=d\omega^\prime+*d\Omega+j_\ell M^{m\ell}*\beta_m\eqno(4.39)$$
where from the continuity equation $\partial_\mu j^\mu=0$
$$*d*\tilde{j}=\nabla_g^2\omega^\prime=-\partial_tj^0~~,~~d*d\Omega=
d\tilde{j}\eqno(4.40)$$
and from the explicit form (2.11) of the particle current
$$j_\ell(t)=\sum_aq_a\dot{r}_a^i(t)\left(\alpha_\ell\right)_i(r_a(t))
\eqno(4.41)$$

Similarily
$$\tilde{\Sigma}=d\Pi^\prime+*d\pi+\Sigma_\ell M^{\ell m}*\alpha_m\eqno(4.42)$$
where from the conservation law $\partial_\mu\Sigma^{\mu\nu}=0$
$$*d*\tilde{\Sigma}=*d*d\Pi^\prime=\partial_t\Sigma_0~~,~~\nabla_g^2\pi=
*d\tilde{\Sigma}\eqno(4.43)$$
where $\Sigma_0=\Sigma_{0i}dx^i$, and the explicit expression (2.12) for the
string current implies
$$\Sigma_\ell(t)=\sum_b\phi_b\int d\sigma\,{\partial X_b^i(t,\sigma)
\over\partial t}{\partial X_b^j(t,\sigma)\over\partial\sigma}
(\beta_\ell)_{ij}(X_b(t,\sigma))\eqno(4.44)$$
As for the gauge fields, we can eliminate the zero modes of the appropriate
Laplacian operators using our previous assumptions and the symmetries
generated by the continuity equations for the sources.

Substituting all of the above Hodge decompositions into (2.16) and integrating
by parts over $\MT$, the Hamiltonian can be written as
$$H=\int_\MT d^3x\,\left(-\theta{\partial j^0\over\partial t}-
K_i{\partial\Sigma^{0i}\over\partial t}-\Omega_i\Phi'^i+
\pi(x)\nabla_g^2\theta'\right)-(a^\ell j_\ell+b^\ell\Sigma_\ell)\eqno(4.45)$$
and in the Schr\"odinger picture we can use the
representations (4.15) and (4.16) to write
$$H=\int_\MT d^3x\,\left[\left(-\theta{\partial j^0\over\partial t}
+{2\pi i\over k}\pi(x){\delta\over\delta\theta}\right)+\left(-K_i
{\partial\Sigma^{0i}\over\partial t}-{2\pi i\over k}\Omega_i P^i_j
{\delta\over\delta K_j}\right)\right]+H_T\eqno(4.46)$$
where the topological Hamiltonian $H_T$, written in terms of the holomorphic
polarization (4.19) and the representation (4.23), is
$$H_T=i(\Sigma_n-\overline\rho_{nm}M^{m\ell}j_{\ell})M^{np}(\Omega^{-1})_{pk}
\gamma^k-{2\pi i\over k}\left(\Sigma_n-\rho_{nm}M^{m\ell}j_{\ell}\right)M^{nk}
{\partial\over\partial\gamma^k}\eqno(4.47)$$
We see that the Hamiltonian separates into three commuting pieces,
one depending only on the local exact part of $A$, one on the local exact
part of $B$, and the other depending only on the global harmonic parts of the
fields. The Schr\"odinger equation (3.7) can therefore be solved by separating
the variables $\theta$, $K$ and $\gamma$ to get the wavefunctions $\Psi_{\rm
phys}[\theta,K,\gamma;t]=\Psi_L[\theta,K;t]\Psi_T(\gamma;t)$.

The local wavefunctionals $\Psi_L$ must solve the local gauge constraints
(2.13), which as in (3.5) are solved in the form
$$\Psi_L[\theta,K;t]=\exp\left[i\int_\MT d^3x\,\left(\theta(x)
j^0(x,t)+K_i(x)\Sigma^{0i}(x,t)\right)\right]\tilde{\Psi}_L(t)\eqno(4.48)$$
and they represent the local gauge symmetries through a 1 dimensional
projective representation as discussed in Section 3.
The remaining piece $\tilde{\Psi}_L$ is found by substituting (4.48) and (4.46)
into the Schr\"odinger equation (3.7), from which we find
$$\tilde{\Psi}_L(t)=\exp\left[{2\pi i\over k}\int_{-\infty}^tdt'\,
\int_\MT d^3x\,\left(\pi(x,t')j^0(x,t')-\Omega_i(x,t')
\Sigma^{0i}(x,t')\right)\right]\eqno(4.49)$$
The evaluation of this integral is done in the Appendix. It is found that
$$\tilde\Psi_L(t)=\exp\left[-{i \over 2k}\sum_{a,b}q_a\phi_b\int_{-\infty}^tdt
^\prime~{d\Phi^{(g)}_{ab}(t^\prime)\over dt^\prime}+{2\pi i\over k}
\int_{-\infty}^tj_\ell(t^\prime)~dt^\prime~M^{m\ell}\int_{-\infty}^{t^\prime}
\Sigma_m(t^{\prime\prime})~dt^{\prime\prime}\right]\eqno(4.50)$$
where $\Phi^{(g)}_{ab}(t)=\Phi^{(g)}(r^i_a(t),X^j_b(t,\sigma))$ is the
generalization of the solid angle (3.11) to a curved space (see the
Appendix, equation (A.20)). Everytime a particle goes around a loop, or a loop
goes around a particle in the opposite direction, $\Phi^{(g)}$ increases by
$4\pi$. Thus we recover the fractional linking of particles and strings found
earlier in Euclidean 3-space. The canonical linking function $\Phi^{(g)}(t)$ is
the adiabatic limit of the standard covariant particle-string linking in
a general 4-manifold $\M4$ [2,3,5].

The solution $\Psi_T$ of the Schr\"odinger equation determined by the
topological Hamiltonian $H_T$ in (4.47) is immediate
$$\eqalign{\Psi_T(\gamma;t)=\exp\biggl[\int_{-\infty}^t&(\Sigma_n(t^\prime)-
\overline\rho_{nm}M^{m\ell}j_\ell(t^\prime))~dt^\prime~M^{np}(\Omega^{-1})
_{pk}\gamma^k\cr&-{2\pi \over k}\int_{-\infty}^t(\Sigma_n(t^\prime)-
\rho_{nm}M^{m\ell}j_\ell(t^\prime))~dt^\prime~M^{np}\cr&\times(\Omega^{-1})_{pq}
M^{kq}\int_{-\infty}^{t^\prime}(\Sigma_k(t^{\prime\prime})-\overline\rho_{kr}
M^{rs}j_s(t^{\prime\prime}))~dt^{\prime\prime}\biggr]\cr&\times\Psi_0^{(q)}
\pmatrix{c\cr d\cr}\left(\gamma^k-{2\pi\over k}M^{nk}\int_{-\infty}^t
(\Sigma_n(t^\prime)-\rho_{nm}M^{m\ell}j_\ell(t^\prime))~dt^\prime \right)\cr}
\eqno(4.51)$$
The last function $\Psi_0$ in (4.51) is at first glance arbitrary since the
combination $\gamma^k-{2\pi\over k}M^{nk}\int_{-\infty}^t
(\Sigma_n(t^\prime)-\rho_{nm}M^{m\ell}j_\ell(t^\prime))~dt^\prime$ in its
argument leads directly to a solution of the equation
$${\partial\Psi_0\over \partial t}=-{2\pi\over k}\left(\Sigma_n-\rho_{nm}M^{m
\ell}j_\ell\right)M^{nk}{\partial\Psi_0\over\partial\gamma^k}\eqno(4.52)$$
which arises from the Schr\"odinger equation after factoring out the
exponential term in (4.51). However, when there are no particles and
strings present we find that $\Psi_T(\gamma)$ must reduce to the coherent
state wavefunctions $\Psi_0(\gamma)$ discussed in Subsection 4.1 above,
which we know to be the functions $\Psi_0^{(q)}$ given by (4.34). (4.51)
determines that part of the full wavefunction which represents the source
currents traversing along the non-trivial homology cycles in $\MT$, as well as
the global gauge symmetries through the multi-dimensional projective
representation discussed in Subsection 4.1.

When we put together $\Psi_L$ and $\Psi_T$ above, use the equation (4.34) for
$\Psi_0^{(q)}$, and substitute in the expressions (2.11) and (2.12) for
the source currents, we obtain the full wavefunction
$$\eqalign{\Psi_{\rm phys}^{(q)}\pmatrix{c\cr d\cr}[\theta,&K,\gamma;t]
=\exp\left[i\sum_aq_a\theta(r_a(t))+i\sum_b\phi_b\int d\sigma~{\partial
X_b^i(t,\sigma)\over\partial\sigma}K_i(X_b(t,\sigma))\right]
\cr&\times\exp\Biggl[-{i\over 2k}\sum_{a,b}q_a\phi_b\left(\Phi^{(g)}_{ab}(t)
-\Phi^{(g)}_{ab}(-\infty)\right)+{k\over4\pi}\gamma^\ell(\Omega^{-1})_{\ell
k}\gamma^k\cr&-i\gamma^k\int_{-\infty}^tj_k(t^\prime)~dt^\prime+{2\pi i
\over k}\int_{-\infty}^tj_\ell(t^\prime)~dt^\prime~M^{m\ell}\int_{-\infty}^t
\Sigma_m(t^\prime)~dt^\prime\cr&-{i\pi\over k}\int_{-\infty}^tj_k(t^\prime)~
dt^\prime~M^{pk}\rho_{pq}M^{q\ell}\int_{-\infty}^tj_\ell(t^\prime)~dt^\prime
\Biggr]\Theta\pmatrix{{c+q\over k_1 k_2M}\cr d\cr}\biggl({Mk_1\over
2\pi}M_{kn}\gamma^k\cr&-k_2\int_{-\infty}^t(\Sigma_n(t^\prime)-\rho_{nm}
M^{m\ell}j_\ell(t^\prime))~dt^\prime\biggm|-k_1k_2M\rho\biggr)\cr}\eqno(4.53)$$
As shown in the Appendix, the wavefunctions (4.53) are independent of the
particular paths of motion of the particles and strings, as long as these
do not intersect. They do, however, depend on their configurations through
the local parts of the gauge fields, the topological current
integrals and the angle function $\Phi^{(g)}_{ab}(t)-\Phi^{(g)}_{ab}(-
\infty)$.

Notice that the states (4.53) also contain $2p$ free parameters $c$ and $d$.
However, it can be observed that
$$\Psi_0^{(q)}\pmatrix{c\cr d\cr}(-\gamma)=(-1)^{4c^\ell d_\ell}
\Psi_0^{(q)}\pmatrix{c\cr d\cr}(\gamma)\eqno(4.54)$$
only when $c^\ell,d_\ell\in\{0,{1\over 2}\}$. For other parameter values
the reflection symmetry $\gamma\to-\gamma$ does not close in the set of
functions (4.34). The values $c^\ell,d_\ell\in\{0,{1\over2}\}$ correspond
to a choice of spin structure on the complex multi-torus formed by the
topological phase space ${\cal P}$ modulo large gauge transformations. This
spin structure increases the number of wavefunctions (4.53) by $4^p$. We have
not found any other symmetry of the theory, for example a symmetry that
relates different complex structures on ${\cal P}$, which would
select special values for $c$ and $d$, and it would be interesting to
investigate this further.

The final step in constructing the Hilbert space of the particle-string
system is to determine the transition amplitudes between physical states.
The inner product in the finite-dimensional vector space spanned by the
wavefunctions (4.53) is given by
$$(\Psi_1,\Psi_2)=\int D\theta~DK~<\Psi_1[\theta,K],\Psi_2[\theta,K]
>_{\cal P}\eqno(4.55)$$
where the inner product on the topological phase space (4.24) is defined by the
usual coherent state measure for the holomorphic polarization (4.19) [23]
$$<\Psi_1,\Psi_2>_{\cal P}~=\int_P\prod_{\ell=1}^pd\gamma^\ell~d\overline{
\gamma}^\ell~\det\Omega^{-1}~\e^{-{k\over2\pi}\gamma^k(\Omega^{-1})_{k\ell}
\overline{\gamma}^\ell}\Psi_1^*(\overline{\gamma})\Psi_2(\gamma)\eqno(4.56)$$
In the subspace spanned by the wavefunctions (4.34), wherein the integrand of
(4.56) is completely invariant under the large gauge transformations (4.20),
the integration in (4.56) can be restricted to the plaquette $P=\{\gamma^\ell=
u^\ell+M^{m\ell}\rho_{mk}v^k:u^\ell,v^\ell\in[0,1]\}$, the multi-torus which is
the reduced phase space of the $\gamma$'s, after dividing out by the volume of
the gauge group in (4.55). With the measure (4.56) and the commutation
relations (4.21), we find that $\gamma^\dagger=\overline{\gamma}$, as it
should be, and also that the infinitesimal variation norm is given by
$\|\delta\gamma\|^2=(\delta\gamma)^k(\Omega^{-1})_{k\ell}(\delta\overline
\gamma)^\ell$. Thus the matrix $\Omega$ given by (4.22) defines a metric on
the topological phase space $\cal P$ of harmonic forms. Notice that the
wavefunctions (4.34) have inner product
$$<\Psi_0^{(q)},\Psi_0^{(q')}>_{\cal P}~=~{\rm det}^{-1/2}\Omega~
\delta^{qq'}\eqno(4.57)$$
and so the basis (4.53) can be used to define an orthonormal basis of the full
physical Hilbert space.

When the current sources are transported around contractable cycles in
$\MT$, there is no additional phase contribution to the wavefunctions (4.53).
This is not true, however, for homologically non-trivial motions of the
particles and strings. Let us consider the effect on the wavefunctions (4.53)
of the motion of particles and strings whereby first the particles wind
$t_k$ times around the $k$-th homology 1-cycle, so that $\int_{-\infty}^{
\tilde t} j_k(t^\prime)~dt^\prime=t_k$ and $\int_{-\infty}^{\tilde t}
\Sigma^k(t^\prime)~dt^\prime=0$, and then afterwards the strings wind
$s_k$ times around the $k$-th homology 2-cycle, so that $\int_{\tilde t}^t
j_k(t^\prime)~dt^\prime=0$ and $\int_{\tilde t}^t\Sigma^k(t^\prime)~dt^
\prime=s_k$. The linkings of these particles and strings
are represented by the angle function $\Phi^{(g)}_{ab}$. Then we find, modulo
these linkings, from (4.33) that the wavefunctions (4.53) are transformed as
$$\eqalign{
\Psi_{\rm phys}^{(q)}\pmatrix{c\cr d\cr}[\theta,K,\gamma;t]&\to\exp
\left[{2\pi i\over k}s_kM^{k\ell}t_\ell+{2\pi i\over k_1M}\left(-s_kc^k+d_k
M^{k\ell}t_\ell\right)-{2\pi i\over k_1M} s_k q^k\right]\cr&\qquad\qquad
\qquad\qquad\times\Psi_{\rm phys}^{(q^k-k_2M^{k\ell}t_\ell)}\pmatrix{c\cr
d\cr}[\theta,K,\gamma;-\infty]\cr&=\sum_{q^\prime} [V(s,t)]_{q q^\prime}
\Psi_{\rm phys}^{(q^\prime)}\pmatrix{c\cr d\cr}[\theta,K,\gamma;-\infty]
\cr}\eqno(4.58)$$
where the matrices
$$[V(s,t)]_{q q^\prime}=\exp\left[{2\pi i\over k}s_kM^{k\ell}t_\ell+{2\pi
i\over k_1M}\left( -s_kc^k+d_kM^{k\ell}t_\ell\right)-{2\pi i\over
k_1M}s_kq^k\right]\delta_{q^k-k_2M^{k\ell}t_\ell,q'^k}\eqno(4.59)$$
form a $(k_1)^p$ dimensional representation of a variation of a clock algebra
$$V(s,t)V(s^\prime,t^\prime)=\e^{{2\pi i\over k}(s_kM^{k\ell}
t_\ell^\prime- s'_kM^{k\ell}t_\ell)}V(s^\prime,t^\prime)V(s,t)\eqno(4.60)$$
which can be viewed as the dual of the algebra (4.27) with $k_1$ and $k_2$
interchanged and the linking matrix $M_{k\ell}$ replaced by its inverse
$M^{\ell k}$.

\subhead{4.3. Discussion and Conclusions}

We have shown in the above that the full Hilbert space of BF theory with
sources is $(k_1k_2M)^p$ dimensional and the wavefunctions carry a
one-dimensional unitary representation of the fractional particle-string
exchange holonomies through the angle function $\Phi^{(g)}_{ab}$ in (4.53),
with statistics parameter ${k\over2\pi}$. The wavefunctions in
addition carry a 1 dimensional projective representation of the local gauge
symmetries, as well as a $(k_2)^p$ dimensional projective representation of the
algebra (4.27) of large gauge transformations with the 2-cocycle
$$\alpha_2(n,m;n',m')={k_1M\over k_2}\left(n^\ell M_{\ell n}m'^n-n'^\ell
M_{\ell n}m^n\right)\eqno(4.61)$$
They also carry the corresponding representation of the dual algebra,
obtained by interchanging $k_1$ and $k_2$ and inverting $M_{k\ell}$, where
the dual tranformations transport the particle and string sources around
the non-trivial homology cycles in $\MT$. What is also interesting is that
when the space $\MT$ does contain non-trivial homology cycles, the spectrum of
the pure source-free BF theory is also non-trivial and provides quantum
representations of the non-trivial cohomology of $\MT$. In all of this, when
the linkings of the homology 1-cycles with the homology 2-cycles is
non-trivial the linking matrix $M^{\ell m}$ plays an important role: It
appears in both the dimensionality of the Hilbert space and the ${\rm U}(1)$
2-cocycle that arises in the projective cohomology
representations of $\MT$. Its appearence is quite nice, since a complete
homological representation should also reflect the non-triviality of
$M^{\ell m}$ in these cases.

The above results provide a full, detailed description of the enlarged
Hilbert space of the particle-string system, with inner product (4.55),
and we see precisely how the fractional linkings of particles and strings
are realized in the physical states. The discrete transformation groups above,
along with the fractional exchange holonomies, are the analogs of the braid
group representations that one obtains in conventional anyon theories [1,24],
i.e. they represent the generators of the fundamental homotopy group of
the particle-string configuration space. For a system of identical, oriented
strings this fundamental group is known as the motion group [15,25]. In
the present case the wavefunctions carry a unitary representation of both the
generators of this group which correspond to the exchange holonomies between
particles and strings and, in the case of a topologically non-trivial
spatial manifold, the generators which are associated with the windings of
the sources around the non-trivial homology cycles of the manifold. These
latter topological generators are dual to the generators of large gauge
transformations, and we see that the BF-theory leads to a duality between
large gauge transformations and linking operations. This inherent duality in
the
wavefunctions yields a quantum representation of the cohomology of the
space on which the BF theory is defined, and also of the possible
non-trivial linkings of the homology cycles. Thus the BF-type
topological field theory description of the
adiabatic transports which occur in such a model is well-suited to
describe the fractional statistics which can occur in 4 dimensions, and
it provides a phenomenological description of anyons propagating in
3-dimensional space, and of the fractional statistics of strings themselves.
It also provides interesting quantum representations of the non-trivial
topological properties that $\MT$ may possess, and of the resulting motion
group on $\MT$.

\head{Appendix: Adiabatic Linking Numbers in Curved Spaces}

In this Appendix we present some details of the calculation of the curved
space solid angle function which was introduced in (4.50).
Many of the calculations in a curved space are performed by decomposing the
space of functions on ${\cal M}_3$ in terms of the eigenfunctions of the
scalar Laplacian operator
$$
\nabla_g^2\psi_\lambda(x)=*d*d\psi_\lambda(x)=\lambda^2\psi_\lambda(x)
\eqno(\A.1)$$
These functions can be normalized so that
$$
\int_{{\cal M}_3}\psi_\lambda*\psi_{\lambda^\prime}=\delta_{\lambda,\lambda
^\prime}
\eqno(\A.2)$$
In the case of degeneracies it is always possible to define an orthonormal set
in the degenerate  subspace, and this will be implicitly assumed in this
Appendix. One important function is the constant function $\psi_0=(\int_{{\cal
M}_3}d^3x)^{-{1\over 2}}$, the single zero
mode solution of (A.1) in the space of functions on ${\cal M}_3$. This
decomposition can be used to represent the Dirac delta function on this space
through the completeness relation $$
\delta^{(3)}(x,y)=\sum_\lambda \psi_\lambda(x)\psi_\lambda(y)~~{\rm or}~~
\delta^{(0,3)}(x,y)d^3y=\sum_\lambda \psi_\lambda(x)*\psi_\lambda(y)
\eqno(\A.3)$$

The above decomposition for functions on ${\cal M}_3$ is the Hodge
decomposition for 0-forms, the elements of $\Lambda^0({\cal M}_3)$. Every
0-form can be expressed in terms of a coexact 0-form, $\psi_\lambda$ for
$\lambda\ne 0$, an exact 0-form, which doesn't exist since there are no
$(-1)$-forms, and a harmonic 0-form, the constant function $\psi_0$. A similar
decomposition exists for 1-forms on ${\cal M}_3$, the elements of
$\Lambda^1({\cal M}_3)$. The exact 1-forms are
$$
\psi^{(e)}_\lambda={d\psi_\lambda \over \sqrt{-\lambda^2}}~~{\rm with}~~
\int_{{\cal M}_3}\psi^{(e)}_\lambda\wedge*\psi^{(e)}_{\lambda^\prime}=\delta_
{\lambda,\lambda^\prime}
\eqno(\A.4)$$
for $\lambda\ne 0$, while the coexact 1-forms are a new set of forms satisfying
$$
*d*\psi^{(c)}_{\tilde\lambda}=0~~,~~(\nabla_g^{(1)})^2\psi^{(c)}_
{\tilde\lambda}=*d*d\psi^{(c)}_{\tilde\lambda}=\tilde\lambda^2\psi^{(c)}_
{\tilde\lambda}\eqno(\A.5)$$
which are also normalized as
$$
\int_{{\cal
M}_3}\psi^{(c)}_{\tilde\lambda}\wedge*\psi^{(c)}_{\tilde\lambda^\prime}=\delta_
{\tilde\lambda,\tilde\lambda^\prime}
\eqno(\A.6)$$
The index $(c)$ will be dropped from here on to simplify notation. We consider
only those $\psi_{\tilde\lambda}$ with $\tilde\lambda\ne 0$,
which in this decomposition then leaves only the harmonic 1-forms given by the
$\alpha_\ell$ introduced at the beginning of Section 4.
They are normalized by the relation (4.3).

It is also possible to define a Dirac delta function for 1-forms in the
following sense. Such a function belongs to the space $\Lambda^1({\cal
M}_3(x))\otimes\Lambda^2({\cal M}_3(y))$, and when it is wedged with a
1-form $\alpha(y)\in\Lambda^1({\cal M}_3(y))$ and the resulting 3-form on
$\Lambda^3({\cal M}_3(y))$ is integrated over ${\cal M}_3$, we are left
with the 1-form $\alpha(x)\in\Lambda^1({\cal M}_3(x))$. From the Hodge
decomposition above, it is rather straightforward to show that this delta
function has the representation in terms of the completeness relation
$$
\delta^{(1,2)}(x,y)=-\sum_{\lambda\ne 0}{d\psi_\lambda(x)*d\psi_\lambda(y)
\over \lambda^2}+\sum_{\tilde\lambda}\psi_{\tilde\lambda}(x)*\psi_
{\tilde\lambda}(y)+\alpha_\ell(x)M^{m\ell}\beta_m(y)\eqno(\A.7)$$
Note that from (A.7) we can represent the linking matrix (4.1) of $\MT$ as
$$ M^{mn}=\int_{\Sigma^m(y)}\int_{L^n(x)}\delta^{(1,2)}(x,y)\eqno(\A.8)$$
which is valid {\it only} with the normalization (4.2) of the harmonic
generators. The Hodge decompositions of 2-forms and 3-forms are the
Hodge-duals of the above decompositions of 1-forms and 0-forms, respectively.

If we look at the definitions of the particle current (2.11) and string
current (2.12) and integrate the time components, we recognize the delta
functions (A.3) and (A.7), respectively. Using the decomposition (4.39)
and (4.40) for the particle current (2.11), we find that
$$
j^0(x,t)=\sum_a q_a\sum_{\lambda}\psi_{\lambda}(x)\psi_{\lambda}(r_a(t))
\eqno(\A.9)$$
$$
\Omega(x,t)=\sum_a q_a\sum_{\tilde\lambda}{1\over\tilde\lambda^2}
*d\psi_{\tilde\lambda}(x){\partial\over \partial t}\left(\int_{r_0}
^{r_a(t)}\psi_{\tilde\lambda}\right)
\eqno(\A.10)$$
and
$$
\omega^\prime(x,t)=-\sum_a q_a\sum_{\lambda\neq0}{1\over\lambda^2}
\psi_{\lambda}(x){\partial\psi_{\lambda}(r_a(t))\over \partial t}=-
{1\over(\nabla_g^2)'}{\partial j^0(x,t)\over \partial t}
\eqno(\A.11)$$
with the harmonic part
$$
j_\ell(t)=\sum_a q_a{\partial \over \partial
t}\left(\int_{r_0}^{r_a(t)}\alpha_\ell\right)
\eqno(\A.12)$$
which is just equation (4.41). Here $r_0$ is some fixed basepoint in $\MT$.

Similarily from the string current decomposition (4.42) and (4.43)
of (2.12) we find
$$
\Sigma_0=\Sigma_{0i}dx^i=\sum_b \phi_b\sum_{\tilde\lambda}\psi_
{\tilde\lambda}(x) \left(\int_{\sigma_b(t)}\psi_{\tilde\lambda}\right)
\eqno(\A.13)$$
$$
\pi(x,t)=-\sum_b \phi_b \sum_{\lambda\ne 0}{1\over\lambda^2}
\psi_{\lambda}(x){\partial\over \partial t}\left(\int_{{\mit\Sigma}_b(t)}
*d\psi_{\lambda}\right)
\eqno(\A.14)$$
and
$$
\Pi'(x,t)=\sum_b \phi_b\sum_{\tilde\lambda}{1\over\tilde{\lambda}^2}
\psi_{\tilde\lambda}(x){\partial\over \partial t}\left(\int_{\sigma_b(t)}
\psi_{\tilde\lambda}\right)={1\over\bigl(\nabla_g^{(1)2}\bigr)'}
{\partial\Sigma_{0i}(x,t)\over \partial t}dx^i
\eqno(\A.15)$$
with the harmonic part
$$
\Sigma_\ell(t)=\sum_b \phi_b{\partial\over \partial t}\left(\int_{{\mit
\Sigma}_b(t)}\beta_\ell\right)\eqno(\A.16)$$
which is equal to (4.44). By $\sigma_b(t)$ we mean the string embedding
$X^i_b(t,\sigma)$, while ${\mit\Sigma}_b(t)$ represents a surface, the string
world sheet projected onto ${\cal M}_3$, having the boundary $\sigma_b(t)$
at time $t$.

Let us now compute the expression (4.49) using the relations (A.9), (A.10),
(A.13) and (A.14). We find that
$$\eqalign{{\cal L}(t)&=
\int_{{\cal M}_3} d^3x~\left(\pi(x,t)j^0(x,t)-\Omega_i(x,t)\Sigma
^{0i}(x,t)\right)\cr&
=\sum_{a,b} q_a \phi_b\left[-\sum_{\lambda\ne0}{1\over\lambda^2}
\psi_{\lambda}(r_a(t)){\partial\over \partial t}\left(\int_{{\mit\Sigma}_b(t)}
*d\psi_{\lambda}\right)-\sum_{\tilde\lambda}\left(
\int_{{\mit\Sigma}_b(t)}*\psi_{\tilde\lambda}\right){\partial\over \partial
t}\left(\int_{r_0} ^{r_a(t)}\psi_{\tilde\lambda}\right)\right]
\cr}\eqno(\A.17)$$
Integrating the first term in (A.17) by parts over time gives
$$\eqalign{{\cal L}(t)
=\sum_{a,b}&q_a \phi_b{\partial\over\partial t}\left[-\sum_{\lambda\ne
0}{1\over\lambda^2}
\psi_{\lambda}(r_a(t))\left(\int_{{\mit\Sigma}_b(t)}*d\psi_{\lambda}\right)
\right]\cr&+\sum_{a,b}q_a\phi_b(\alpha_\ell)_i(r_a(t))\dot{r}_a^i(t)M^{m\ell}
\left(\int_{{\mit\Sigma}_b(t)}\beta_m\right)
\cr&-\sum_{a,b} q_a \phi_b\Biggl[-\sum_{\lambda\ne 0}{1\over\lambda^2}
\left(\int_{{\mit\Sigma}_b(t)}*d\psi_\lambda\right)\partial_i\psi_\lambda(
r_a(t))\cr&+\sum_{\tilde\lambda}\left(\int_{{\mit\Sigma}_b(t)}*\psi_{\tilde
\lambda}\right)\left(\psi_{\tilde
\lambda}\right)_i(r_a(t))+\left(\int_{{\mit\Sigma}_b(t)}\beta_m\right)
M^{m\ell}
\left(\alpha_\ell\right)_i(r_a(t))\Biggr]\dot{r}_a^i(t)\cr}\eqno(\A.18)$$
The last 3 terms in (A.18) can be collected together to give the delta
function (A.7) integrated over ${\mit\Sigma}_b(t)$, and then comparing the
second term in (A.18) with (A.12) and (A.16) we find
$${\cal L}(t)
=-{1\over 4\pi}\sum_{a,b} q_a \phi_b {d\Phi^{(g)}_{ab}(t)\over
dt}+j_\ell(t)M^{m\ell}\int_{-\infty}^t\Sigma_m(t^\prime)~dt^\prime
\eqno(\A.19)$$
where we have defined the function $\Phi^{(g)}_{ab}$ by
$$\eqalign{
\Phi^{(g)}_{ab}(t)=4\pi\int_{-\infty}^t\biggl(\int_{{\mit\Sigma}_b(t^\prime)}&
\delta^{(1,2)}(r_a(t^\prime),X_b(t',\sigma))\biggr)_i\dot{r}_a^i(t^\prime)
{}~dt^\prime\cr&+4\pi\sum_{\lambda\ne 0}{1\over\lambda^2}\psi_
\lambda(r_a(t))\left(\int_{{\mit\Sigma}_b(t)}*d\psi_\lambda\right)
\cr}\eqno(\A.20)$$

We need to check that $\Phi^{(g)}_{ab}$ is independent of the chosen
surface ${\mit\Sigma}_b(t)$ given
its boundary $\sigma_b(t)$. If we choose a different surface $\tilde
{\mit\Sigma}_b(t)$ with ${\mit\Sigma}_b-\tilde{\mit\Sigma}_b=\partial B_b$ for
some volume $B_b$, then the second term in (A.20) will change by
$$
\delta\Phi^{(g)}_{ab}(t)=4\pi\sum_{\lambda\ne 0}{1\over\lambda^2}\psi_
\lambda(r_a(t))\left(\int_{{\mit\Sigma}_b(t)-\tilde{\mit\Sigma}_b(t)}*d\psi_
\lambda\right)\eqno(\A.21)$$
Using Stokes' theorem we then find
$$\delta\Phi^{(g)}_{ab}(t)
=4\pi\sum_{\lambda\ne 0}\int_{B_b(t)}\psi_\lambda(r_a(t))\psi_\lambda(x)~d^3x
=4\pi\int_{B_b(t)}\left[\delta^{(3)}(r_a(t),x)-\psi_0\psi_0\right]~d^3x
\eqno(\A.22)$$
This shows that if we continuously deform ${\mit\Sigma}_b(t)$ then the second
term in (A.20) does not change unless we cross the particle at $r_a(t)$, and
then the change is $4\pi$ which will be cancelled by the first term in (A.20).
The contribution from the $\psi_0$ function in (A.22) will vanish when we
sum over $q_a$ and $\phi_b$ in (A.19).

A string encircling a fixed particle (i.e. $r_a(t)$ constant) once can be
represented by ${\mit\Sigma}_b(t_1)={\mit\Sigma}_b(t_2)$ after it has swept out
a closed volume containing the particle in time $t_2-t_1$, and in this case
the second term in (A.20) will add $4\pi$ to $\Phi^{(g)}_{ab}$. Alternatively,
if the string is fixed we can choose ${\mit\Sigma}_b(t)$ constant, and when
a particle encircles the string and returns back to its original position the
only contribution to $\Phi^{(g)}_{ab}$ is the first term in (A.20) which
precisely counts the number of times it crosses ${\mit\Sigma}_b$, giving again
the correct linking number. The function (A.20) therefore has the properties
that it increases by $4\pi$ everytime that particle $a$ and string $b$ link
themselves exactly once, and furthermore it is independent of the paths of
the particles and strings, as long as they do not intersect. (A.20) also
reduces to the standard Euclidean space solid angle when the string and
particle paths are infinitesimal, since then this function becomes the
angle function (3.11). The canonical linking function $\Phi^{(g)}(t)$ above
is the adiabatic limit of the covariant linking number
$$I_{\M4}({\mit\Sigma},L)=\int_{B({\mit\Sigma})}\Delta_L\eqno(\A.23)$$
in $\M4$ of a loop $L$ and a closed surface $\mit\Sigma$ [2,3]. Here
$B({\mit\Sigma})$ is a volume bounded by $\mit\Sigma$ and $\Delta_L$ is the
deRham current [17] which is the Poincar\'e-dual of the embedding of $L$ in
$\M4$. The deRham current can be written as $\Delta_L=\int_L\Delta$
where $\Delta$ is the Dirac delta function $\delta^{(1,3)}(r(\tau),x)$ in the
space $\Lambda^1(\M4(r(\tau)))\otimes\Lambda^3(\M4(x))$ which restricts the
domain of integration over $\M4$ to $L$, i.e. $\int_{\M4}\Delta_L\wedge\alpha
=\int_L\alpha$ for any 1-form $\alpha$.

\vfill\eject

\singlespace

\centerline{\bf References}

\bigskip

\item{[1]} G. W. Semenoff, in Proceedings of the Karpacz Winter School in
Theoretical Physics (1993).
\line{\hfill}
\item{[2]} D. Birmingham, M. Blau, M. Rakowski and G. Thompson, Phys. Rep.
{\bf 209} (1991), 129.
\line{\hfill}
\item{[3]} M. Blau and G. Thompson, Ann. Phys. {\bf 205} (1991), 130.
\line{\hfill}
\item{[4]} G. T. Horowitz, Commun. Math. Phys. {\bf 125} (1989), 417; M. Blau
and G. Thompson, Phys. Lett. {\bf B228} (1989), 64;
A. Lahiri, Mod. Phys. Lett. {\bf A8} (1993), 2403.
\line{\hfill}
\item{[5]} G. T. Horowitz and M. Srednicki, Commun. Math. Phys. {\bf 130}
(1990), 83; I. Oda and S. Yahikozawa, Phys. Lett. {\bf B238} (1990), 272; S.
Wu, Commun. Math. Phys. {\bf 136} (1991), 157.
\line{\hfill}
\item{[6]} P. G. de Gennes, {\it Superconductivity of Metals and Alloys},
W. A. Benjamin, Inc. (New York) (1966); A. P. Balachandran, V. P. Nair,
B.-S. Skagerstam and A. Stern, Phys. Rev. {\bf D26} (1982), 1443; T. R.
Govindarajan, J. Phys. {\bf G8} (1982), L17.
\line{\hfill}
\item{[7]} C. P. Burgess and A. Kshirsagar, Nucl. Phys. {\bf B324} (1989),
157; J. M. Molera and B. A. Ovrut, Phys. Rev. {\bf D40} (1989), 1146; K. Lee:
The Dual Formulation of Cosmic Strings and Vortices, CERN preprint
CERN-TH-6780/93 (1993).
\line{\hfill}
\item{[8]} M. J. Bowick, S. B. Giddings, J. A. Harvey, G. T. Horowitz and
A. Strominger, Phys. Rev. Lett. {\bf 61} (1988), 2823; T. J. Allen, M. J.
Bowick and A. Lahiri, Phys. Lett. {\bf B237} (1989), 47; B. A. Campbell,
M. J. Duncan, N. Kaloper and K. A. Olive, Phys. Lett. {\bf B251} (1990), 34;
Nucl. Phys. {\bf B351} (1991), 778; S. Coleman, J. Preskill and F. Wilczek,
Nucl. Phys. {\bf B378} (1992), 175.
\line{\hfill}
\item{[9]} M. G. Alford and F. Wilczek, Phys. Rev. Lett. {\bf 62} (1989),
1071; L. M. Krauss and F. Wilczek, Phys. Rev. Lett. {\bf 62} (1989), 1221;
M. G. Alford, J. March-Russel and F. Wilczek, Nucl. Phys. {\bf B337} (1990),
695; J. Preskill and L. M. Krauss, Nucl. Phys. {\bf B341} (1990), 50.
\line{\hfill}
\item{[10]} M. I. Polikarpov, U.-J. Wiese and M. A. Zubkov: String
Representation of the Abelian Higgs Theory and Aharanov-Bohm Effect on the
Lattice, Institute of Theoretical and Experimental Physics preprint
ITEP-16-1993 (1993).
\line{\hfill}
\item{[11]} A. S. Schwarz, Commun. Math. Phys. {\bf 67} (1979), 1; J.
Gegenberg and G. Kunstatter, Phys. Lett. {\bf B321} (1994), 193.
\line{\hfill}
\item{[12]} X. Fustero, R. Gambini and A. Trias, Phys. Rev. Lett. {\bf 62}
(1989), 1964; J. A. Harvey and J. Liu, Phys. Lett. {\bf B240} (1990), 369.
\line{\hfill}
\item{[13]} E. Witten, Phys. Lett. {\bf B149} (1984), 351.
\line{\hfill}
\item{[14]} M. Kalb and P. Ramond, Phys. Rev. {\bf D9} (1974), 2273; R. Rohm
and E. Witten, Ann. Phys. {\bf 170} (1986), 454.
\line{\hfill}
\item{[15]} C. Aneziris, A. P. Balachandran, L. H. Kauffman and A. M.
Srivastava, Int. J. Mod. Phys. {\bf A6} (1991), 2519.
\line{\hfill}
\item{[16]} A. P. Balachandran, G. Bimonte and P. Teotonio-Sobrinho, Mod.
Phys. Lett. {\bf A8} (1993), 1305; A. P. Balachandran and P.
Teotonio-Sobrinho,
Int. J. Mod. Phys. {\bf A8} (1993), 723; {\bf A9} (1994), 1569.
\line{\hfill}
\item{[17]} R. Bott and L. W. Tu, {\it Differential Forms in Algebraic
Topology}, Springer-Verlag (New York) (1986).
\line{\hfill}
\item{[18]} D. B. Ray and I. M. Singer, Adv. Math. {\bf 7} (1971), 145.
\line{\hfill}
\item{[19]} H. Flanders, {\it Differential Forms with Applications to the
Physical Sciences}, Dover Publications (New York) (1989).
\line{\hfill}
\item{[20]} O. Alvarez, Commun. Math. Phys. {\bf 100} (1985), 279; A. P.
Polychronakos, Nucl. Phys. {\bf B281} (1987), 241.
\line{\hfill}
\item{[21]} E. Witten, Commun. Math. Phys. {\bf 121} (1989), 351.
\line{\hfill}
\item{[22]} D. Mumford, {\it Tata Lectures on Theta}, Birkhauser (Basel)
(1983).
\line{\hfill}
\item{[23]} A. M. Perelomov, {\it Generalized Coherent States and Their
Applications}, Springer-Verlag (Berlin) (1986).
\line{\hfill}
\item{[24]} M. Bergeron and G. W. Semenoff, Ann. Phys. (1994), in press.
\line{\hfill}
\item{[25]} D. L. Goldsmith, Mich. Math. J. {\bf 28} (1981), 3.

\end